%% file: map-det_YRM5.tex
\documentclass[a4paper]{jpconf}
\usepackage[english]{babel}
\usepackage[latin1]{inputenc}

\usepackage{hyperref}
\usepackage{amsmath, amssymb, amsfonts, mathrsfs}
\usepackage{xcolor}
\usepackage{exscale}
\usepackage{graphicx}
\usepackage{latexsym}
\graphicspath{{images/}}

\usepackage{dsfont}

\newcommand{\proj}[1]{|#1\rangle\langle#1|}
\DeclareMathOperator{\Id}{\mathds{1}}

\newcommand{\map}[1]{\mathscr{#1}}

\input{myqcircuit}

\begin{document}

\title{Detection methods to rule out completely co-positive and bi-entangling operations}

\author{C. Macchiavello and M. Rossi$^\dagger$}
\address{Dipartimento di Fisica and INFN-Sezione di Pavia, 
via Bassi 6, 27100 Pavia, Italy}
\ead{$^\dagger$ matteo.rossi@unipv.it}

\date{\today}
\begin{abstract}
In this work we extend the quantum channel detection method developed in Refs. \cite{ns1, ns2} in order to detect other interesting convex sets of quantum channels. First we work out a procedure to detect non completely co-positive maps. Then we focus on the set of so-called bi-entangling operations and show how a map outside this set can be revealed. 
In both cases we provide explicit examples showing the theoretical 
technique and the corresponding experimental procedure.
\end{abstract}

\section{Introduction}

In quantum information it is of great importance to characterise 
quantum communication channels or quantum devices without necessarily
performing quantum process tomography. Actually, quantum process 
tomography requires a large number of experimental resources, while one is 
usually interested in few properties of the quantum channel under 
consideration, as e.g. whether the channel has some entangling power. 
In many realistic implementations some a priori information on the form of 
the channel is available, hence, the quantum channel detection (QCD) method 
developed in \cite{ns1, ns2} can be applied. Besides being less informative 
than the full quantum process tomography, the QCD method allows us to test 
the property of interest with a much smaller experimental effort.

In this work we will discuss in detail how to detect two sets of quantum 
channels, namely quantum channels that are not completely co-positive (CCOP)
and the set of operations that are not bi-entangling (BE). Both sets are of 
great interest as they are connected via the Choi-Jamiolkowski isomorphism to 
PPT states\footnote{A state of a bipartite system is PPT if the partial transpose of its density matrix is positive semi-definite, otherwise it is NPT.} \cite{copositive} and to the problem of classical simulatability 
of quantum computation \cite{virmani}, respectively.

This work is organized as follows. In Sec. \ref{s:idea} we will review the  
main idea of QCD following Refs. \cite{ns1,ns2}. 
In Sec. \ref{s:co-pos} we will discuss a method to detect non CCOP maps. 
In Sec. \ref{s:bi-ent} we will study how to reveal quantum channels that are 
not BE operations, and we finally summarize the main results in Sec. 
\ref{s:conc}.

\section{The general QCD method}
\label{s:idea}

The QCD method proposed in Refs. \cite{ns1,ns2} relies on the concept 
of witness operators \cite{horo-ter} and the Choi-Jamiolkowski isomorphism 
\cite{jam}. We briefly remind both of them in the following. 

A state $\rho$ is entangled if and only if there exists a hermitian operator 
$W$ such that $\Tr[W\rho]< 0$ and $\Tr[W\rho_{sep}]\geq 0$ for all separable 
states; such an operator is called an entanglement witness. The 
Choi-Jamiolkowski isomorphism provides a one-to-one correspondence between 
completely positive (CP) maps $\map{M}$ acting on 
$\mathcal{D(H)}$ (the set of density operators on $\mathcal{H}$, 
with finite dimension $d$) and positive operators $C_{\map{M}}$ on 
$\mathcal{H_A\otimes H_B}$ 
(named Choi states), where $A$ and $B$ here denote the two subsystems
on which the Choi state is defined. The isomorphism can be stated as 
\begin{equation}
\map{M}\Longleftrightarrow C_{\map{M}}=(\map{M}\otimes\map{I})[\proj{\alpha}],
\end{equation}
where $\map{I}$ is the identity map, and $\ket{\alpha}$ is the maximally entangled state with respect to the bipartite space $\mathcal{H\otimes H}$, i.e. $\ket{\alpha}=\frac{1}{\sqrt d}\sum_{k=1}^d\ket{k}\ket{k}$. The above isomorphism can be exploited to link convex sets of quantum channels to particular convex 
sets of quantum states. 

As a simple example consider the convex set of entanglement breaking (EB) 
channels. A channel $\map{E}$ is EB if and only if its Choi state 
$C_\map{E}$ is separable \cite{EB}. Therefore, the detection of entanglement 
of $C_\map{E}$ by means of a suitable witness operator $W_{EB}$ implies 
that the implemented quantum channel $\map{E}$ is not EB \cite{ns1}.

Although the general QCD method applies to several convex sets of quantum 
channels, as e.g. EB and separable maps, in the following it 
will be explicitly studied for the convex sets of CCOP channels and 
BE operations.

\section{Completely co-positive channels}
\label{s:co-pos}

In this section we will consider the set of CCOP channels. 
A CP map $\map{C}$ acting on a qudit ($d$-dimensional system) is CCOP if and 
only if the composite map $\map{C_T}=\map{T}\circ\map{C}$, where $\map{T}$ 
is the transposition map, is CP. Since a quantum map is CP if and only if the 
corresponding Choi operator is positive, we can restate the above definition 
as follows: a CP map $\map C$ is CCOP if and only if the Choi operator 
$C_{\map{C_T}}$ related to the composite map $\map{C_T}$ is positive. 

By the above correspondence we will develop a method to detect 
whether a map is non CCOP by adapting techniques developed for the 
detection of  non positive partial transposed (NPT) entangled states 
\cite{copositive}. Consider then a map $\map{M}$ that does not belong to the 
set of CCOP channels. From the above definition it follows that the bipartite 
Choi state $C_{\map{M_T}}=(\map{T}_A\otimes\map{I})[C_\map{M}]$ has at least 
one negative eigenvalue. Let $\lambda_-$ be the most negative eigenvalue 
corresponding to the eigenvector $\ket{\lambda_-}$. 
The following operator, i.e. 
\begin{equation}\label{ccop}
W_{\text{CCOP}}=\proj{\lambda_{-}}^{\map{T}_A},
\end{equation}
is thus suitable to detect the NPT state $C_{\map{M_T}}$ corresponding to the 
non CCOP map ${\map{M_T}}$. Notice that the transposition map on the Choi 
state acts only on the first qudit, i.e.  $\map{T}_A$. 

As an illustrative example we consider the case of the dephasing noise $\map{D}$ acting on a single qubit, defined by the following trace preserving CP map
\begin{equation}
\map{D}[\rho]=p\rho+(1-p)\sigma_z \rho \sigma_z\;,
\end{equation}
where $\sigma_z$ is a Pauli operator\footnote{Hence, dephasing noise consists in either leaving the input state unchanged (with probability $p$) or applying a phase flip $\sigma_z$ (with $1-p$). Generally speaking, it represents a loss 
of quantum coherence in the off-diagonal terms of the regarded system.}.
It is easy to see that the Choi state $C_\map{D}$ corresponding to $\map{D}$ 
takes the form
\begin{equation}
C_\map{D}=p\proj{\alpha}+(1-p)\proj{\phi^-},
\end{equation}
with $\ket{\phi^-}=\frac{1}{\sqrt 2}(\ket{00}-\ket{11})$. The above state can 
be shown to be NPT whenever $p\neq 1/2$. It is then possible to derive the 
following detection operator \cite{ent-wit, jmo} from Eq. \ref{ccop}:
\begin{equation}
W_{\text{CCOP},\map{D}}=
\begin{cases}
\frac{1}{4}(\Id\otimes\Id +\sigma_x\otimes\sigma_x
+\sigma_y\otimes\sigma_y-\sigma_z\otimes\sigma_z) & \text{ for } p<\frac{1}{2},\\
\frac{1}{4}(\Id\otimes\Id -\sigma_x\otimes\sigma_x
-\sigma_y\otimes\sigma_y -\sigma_z\otimes\sigma_z) & \text{ for } p>\frac{1}{2}.\\
\end{cases}
\end{equation}

\begin{figure}[t!]
\begin{equation*}
\Qcircuit
@C=1em @R=.7em 
{
\multiprepareC{1}{\ket{\alpha}} & \gate{\map{D}} &\ustick{A}\qw &\meter &\pureghost{AAAA}\\
 \pureghost{\ket{\alpha}} & \qw & \ustick{B }\qw &\meter  &\ustick{W_{\text{CCOP},\map{D}}}  
 \gategroup{1}{4}{2}{5}{.3em}{--}
}
\end{equation*}
\caption{Experimental scheme showing the detection of the dephasing channel 
$\map D$ as a non CCOP channel. Notice that the expectation value of the witness $W_{\text{CCOP},\map{D}}$, namely $\Tr[W_{\text{CCOP},\map{D}}C_{\map{D}}]$, can be measured locally.}
\label{fig1}
\end{figure}
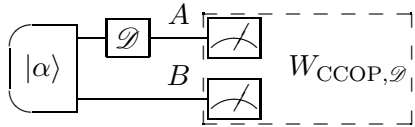

This method can be experimentally implemented by preparing a two-qubit state
in the maximally entangled state  $\ket{\alpha}$, then operating with the 
quantum channel $\map{D}$ to be detected on one of the two qubits and measuring the 
operator $W_{\text{CCOP},\map{D}}$ acting on both qubits at the end (see Fig. \ref{fig1}).
If the resulting average value $\Tr[W_{\text{CCOP},\map{D}}C_{\map{D}}]$ is negative, we can then conclude that the Choi state $C_\map{D_T}=\map{T}\circ \map{D}$ is NPT and that the channel under consideration is not CCOP.

Finally, we would like to stress that, since every NPT state is entangled, 
the detection of a non CCOP channel $\map{M}$ implies that $\map{M}$ is 
not EB as well, but the opposite does not hold in general. 
Actually the set of EB channels is in general a subset of the 
CCOP channels. The two sets coincide only when the channels act on 
two-dimensional systems. 
From the perspective of QCD, this implies that for higher dimensional systems
a quantum channel which is detected as non EB may nevertheless belong to the 
set of CCOP maps.

\section{Bi-entangling operations}
\label{s:bi-ent}

In this section we will focus on BE operations, a class of quantum channels 
that can generate at most bipartite entanglement. They were introduced in 
Ref. \cite{virmani} in the context of quantum computation and were shown to 
be efficiently simulatable classically. BE operations are quantum channels 
acting on bipartite systems $AB$ (of finite dimension $d$) in such a way that they can
 be expressed
as convex combinations of (a) separable operations, (b) operations that
swap the two qudits and then act as a separable operation,
and (c) EB channels, that break any entanglement between the two qudits 
on which the channel acts and extra ancillae \cite{virmani}. 
Via the Choi-Jamiolkowski isomorphism we can then characterize the set of BE 
operations in terms of the corresponding Choi states. 

Consider a BE operation $\map{M}_{BE}$ acting on the bipartite system $AB$.
The Choi state $C_{\map{M}_{BE}}$ associated to $\map{M}_{BE}$ is then a 
four-partite state (composed of subsystems $A$, $B$, $C$ and $D$). 
Separable channels have separable Choi states with respect to the bipartition 
$AC|BD$ \cite{kraus_sep}. As a consequence, channels of type (b), with a swap 
gate followed by separable channels, have separable Choi states in $AD|BC$. 
EB channels correspond to separable Choi states in the bipartition $AB|CD$.
A general Choi state $C_{\map{M}_{BE}}$ for a BE channel
can then be written as a convex combination of four-partite states 
biseparable with respect to bipartitions $AC|BD$, $AD|BC$ 
and $AB|CD$, namely
\begin{equation}
C_{\map{M}_{BE}}=
		p\sum_i p_i C_i^{(AC|BD)}+q\sum_j q_j C_j^{(AD|BC)}+r\sum_k r_k C_k^{(AB|CD)},
\end{equation}
where $(p,q,r)$, $\{p_i\}$, $\{q_j\}$ and $\{r_k\}$ are probability 
distributions. Notice that the first term corresponds to the set (a), 
the second to (b) and the third to (c). 
In other words, the Choi states $C_{\map{M}_{BE}}$ corresponding to BE 
operations lie in the convex hull of states biseparable with respect to 
the bipartitions $AC|BD$, $AD|BC$ 
and $AB|CD$ for the four-partite system $ABCD$. 
We name this convex set of four-partite Choi states corresponding to 
BE operations as $S_{BE}$. It is now possible to develop detection procedures 
for BE operations by employing suitable
witness operators that detect the corresponding Choi state with
respect to the biseparable states belonging to $S_{BE}$. 

We will now focus on the case of a unitary transformation $U$
acting on two $d$-dimensional systems. The corresponding Choi state is pure 
and given by $\ket{U}=(U\otimes\Id)\ket{\alpha}$.
Therefore a suitable detection operator for $U$ as a non BE operation
can be constructed as \cite{ns1,ns2}
\begin{equation}\label{W}
W_{BE,U}=\alpha_{BE}^2\Id-C_U\;,
\end{equation}
where $C_U=\ket{U}\bra{U}$, and the coefficient $\alpha_{BE}$ is the overlap 
between the closest biseparable state in the set $S_{BE}$ and the entangled 
state $\ket{U}$, namely
\begin{equation}
\alpha_{BE}^2 =\max_{\map{M}_{BE}}\bra{U}C_{\map{M}_{BE}}\ket{U}.
\end{equation}
Since the maximum of a linear function over a convex set is 
always achieved on the extremal points, the maximum involved in $\alpha_{BE}$ 
can be always 
calculated by maximising over the pure biseparable states in $S_{BE}$, i.e.
\begin{equation}
\alpha_{BE} =\max_{\ket{\Xi}\in S_{BE}}|\bra{\Xi}U\rangle |.
\end{equation}
By exploiting the Schmidt decomposition \cite{nc} of the state $\ket{U}$, the 
maximization above can be expressed analytically as
\begin{equation}
\alpha_{BE} =\max_i\max_\lambda \lambda_i(U),
\end{equation}
where the index $i$ labels the bipartitions $AC|BD$, $AD|BC$ and $AB|CD$, 
and $\lambda_i(U)$ are the Schmidt coefficients of $\ket{U}$ in the 
bipartition $i$. Therefore, in order to find the coefficient $\alpha_{BE}$ 
one has to find the maximal Schmidt coefficient of  $\ket{U}$ for a fixed 
bipartite splitting and then maximize it among all the bipartitions 
involving only two versus two subsystems.

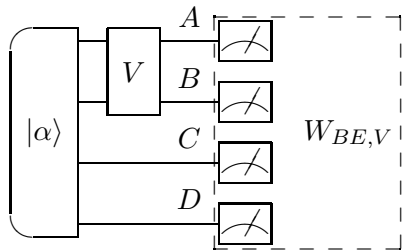
\begin{figure}[t!]
\begin{equation*}
\Qcircuit
@C=1em @R=.7em 
{
\multiprepareC{3}{\ket{\alpha}} & \multigate{1}{V} & \ustick{A} \qw &  \meter & \pureghost{AAA}\\
\pureghost{\ket{\alpha}} 		& \ghost{V}  	& \ustick{B} \qw &  \meter & \pureghost{AAA}\\
\pureghost{\ket{\alpha}} 		& \qw	 & \ustick{C} \qw &  \meter &\ustick{W_{BE,V}}  \\
\pureghost{\ket{\alpha}} 		& \qw	& \ustick{D} \qw &  \meter & \pureghost{AAA}
\gategroup{1}{4}{4}{5}{.3em}{--}
}
\end{equation*}
\caption{Experimental scheme implementing the detection of the gate $V$ defined in Eq. \ref{Wdef} as a non BE operation.}
\label{fig2}
\end{figure}

As an example of the above procedure consider the following unitary operation 
$V$ acting on a two-qubit system
\begin{equation}\label{Wdef}
V=
\begin{pmatrix}
1 & 0 & 0 & 0 \\
0 & 0 & 1 & 0 \\
0 & 1 & 0 & 0 \\
0 & 0 & 0 & -1
\end{pmatrix}\;
\end{equation}
The gate $V$ is a modified swap gate such that it is no longer a BE operation. 
The coefficient $\alpha_{BE}$ for $V$ can be computed following the steps 
outlined above. The Choi state $\ket{V}$ associated to the gate $V$ is
given by
\begin{equation}
\ket{V}=\frac{1}{\sqrt 2}(\ket{\alpha}_{AD}\ket{00}_{BC}+\ket{\phi^-}_{AD}\ket{11}_{BC}),
\end{equation}
and the Schmidt coefficients of $V$ with respect to the bipartitions $AC|BD$, $AD|BC$ and $AB|CD$ can be easily computed as $\lambda_{AC|BD}(V)=(\frac{1}{2},\frac{1}{2},\frac{1}{2},\frac{1}{2})$, $\lambda_{AD|BC}(V)=(\frac{1}{\sqrt 2},\frac{1}{\sqrt 2},0,0)$ and $\lambda_{AB|CD}(V)=(\frac{1}{2},\frac{1}{2},\frac{1}{2},\frac{1}{2})$. Therefore, the coefficient $\alpha_{BE}$ equals $1/\sqrt 2$ and a suitable 
detection operator in order to detect $V$ as a non BE operation takes the form
\begin{equation}\label{WV}
W_{BE,V}=\frac{1}{2}\Id - C_V.
\end{equation}
From an experimental point of view, the detection procedure can be implemented
as follows: prepare a four-partite qubit system in the state  
$\ket{\alpha}=\ket{\alpha}_{AC}\ket{\alpha}_{BD}$, apply 
the quantum gate $V$ to qubits A and B, and finally perform a suitable set of local measurements in order to measure the operator \eqref{WV}. 
If the resulting average value $\Tr[W_{BE,V}C_V]$ is negative then the quantum channel is 
detected as a non BE operation.
The experimental scheme is shown in Fig. \ref{fig2}.  

We conclude this section by noticing that the method described above leads to different 
detection operators with respect to the detection method for non separable maps \cite{ns1}. 
Indeed, already in the case of two qubits, the optimal witness operator that 
detects the gate $V$ as a non separable channel is \cite{ns1}
\begin{equation}
W_{Sep,V}=\frac{1}{4}\Id - C_V.
\end{equation}
As expected, the detection operator above is weaker than $W_{BE,V}$ in Eq. 
\ref{WV} in the sense that it leads to a negative expectation value for a 
smaller set of CP maps. This is due to the fact that BE maps are a strict 
subset of separable maps, and actually the set of separable Choi 
states in the bipartition $AC|BD$, corresponding to separable maps, is a 
strict subset of $S_{BE}$.

\section{Conclusions}
\label{s:conc}

In summary, after a brief review of the general quantum channel detection 
method proposed in \cite{ns1}, based on the Choi-Jamiolkowski isomorphism and witness operators, we have developed a method to detect maps 
that do not belong to specific convex sets, i.e. the completely 
co-positive maps and the bi-entangling operations. 
Significant examples of non co-positive and non bi-entangling operations have 
been considered in detail, showing both the underlying theoretical techniques 
and the corresponding experimental schemes. 
We stress that the method works when some a priori knowledge on the quantum 
channel is available, it requires fewer measurements than quantum process 
tomography and it is achievable experimentally with present day 
technology \cite{exp}. In particular, the detection method for non 
entanglement 
breaking channels and for non separable maps has already been demonstrated in
a quantum optical experiment \cite{exp2}.

\section*{Acknowledgements}

We thank Shashank Virmani for fruitful suggestions.

\end{document}

%% file: myqcircuit.tex
\usepackage[matrix,frame,arrow]{xy}
\usepackage{amsmath}
\newcommand{\bra}[1]{\left\langle{#1}\right\vert}
\newcommand{\ket}[1]{\left\vert{#1}\right\rangle}
\newcommand{\qw}[1][-1]{\ar @{-} [0,#1]}

\newcommand{\gate}[1]{*{\xy *+<.6em>{#1};p\save+LU;+RU **\dir{-}\restore\save+RU;+RD **\dir{-}\restore\save+RD;+LD **\dir{-}\restore\POS+LD;+LU **\dir{-}\endxy} \qw}
\newcommand{\meter}{\gate{\xy *!<0em,1.1em>h\cir<1.1em>{ur_dr},!U-<0em,.4em>;p+<.5em,.9em> **h\dir{-} \POS <-.6em,.4em> *{},<.6em,-.4em> *{} \endxy}}

\newcommand{\multigate}[2]{*+<1em,.9em>{\hphantom{#2}} \qw \POS[0,0].[#1,0];p !C *{#2},p \save+LU;+RU **\dir{-}\restore\save+RU;+RD **\dir{-}\restore\save+RD;+LD **\dir{-}\restore\save+LD;+LU **\dir{-}\restore}
\newcommand{\ghost}[1]{*+<1em,.9em>{\hphantom{#1}} \qw}

\newcommand{\gategroup}[6]{\POS"#1,#2"."#3,#2"."#1,#4"."#3,#4"!C*+<#5>\frm{#6}}

\newcommand{\ustick}[1]{*!D!<0em,-.5em>=<0em>{#1}}

\newcommand{\Qcircuit}[1][0em]{\xymatrix @*=<#1>}

\newcommand{\pureghost}[1]{*+<1em,.9em>{\hphantom{#1}}}
\newcommand{\multiprepareC}[2]{*+<1em,.9em>{\hphantom{#2}}\save[0,0].[#1,0];p\save !C
  *{#2},p+RU+<0em,0em>;+LU+<+.8em,0em> **\dir{-}\restore\save +RD;+RU **\dir{-}\restore\save
  +RD;+LD+<.8em,0em> **\dir{-} \restore\save +LD+<0em,.8em>;+LU-<0em,.8em> **\dir{-} \restore \POS
  !UL*!UL{\cir<.9em>{u_r}};!DL*!DL{\cir<.9em>{l_u}}\restore}
